\documentclass[]{aa}
\usepackage{psfig}
\begin{document}

\thesaurus{06(08.14.2; 08.09.2 CI Aql)}

\title{The 2000 outburst of the recurrent nova\\
CI~Aquilae: optical spectroscopy\thanks{Based on the data
obtained at the David Dunlap Observatory, University of Toronto}}

\author{L.L. Kiss\inst{1,4} \and J.R. Thomson\inst{2} \and
W. Ogloza\inst{3} \and G. F\H{u}r\'esz\inst{1,4} \and K. Szil\'adi\inst{1,4}}

\institute{Department of Experimental Physics and Astronomical Observatory,
University of Szeged,
Szeged, D\'om t\'er 9., H-6720 Hungary \and
University of Toronto, David Dunlap Observatory, Richmond Hill, Canada \and
Mt. Suhora Observatory of the Pedagogical University, Cracow, Poland \and
Guest Observer at Konkoly Observatory}

\titlerunning{The 2000 outburst of CI~Aquilae}
\authorrunning{Kiss et al.}
\offprints{l.kiss@physx.u-szeged.hu}
\date{}

\maketitle
 
\begin{abstract}

We present low- and medium resolution spectra of the recurrent
nova CI~Aquilae taken at 14 epochs in May and June, 2000.
The overall appearance is similar to other U~Sco-type recurrent
novae (U~Sco, V394 CrA). Medium resolution ($\lambda/\Delta\lambda\approx7000-10000$)
hydrogen and iron profiles suggest
an early expansion velocity of 2000--2500 km~s$^{-1}$.
The H$\alpha$ evolution is followed from $\Delta$t = $-$0.6 d to
+53 d, starting from a nearly Gaussian shape to a double peaked
profile through strong P-Cyg profiles. The interstellar component
of the sodium D line and two diffuse interstellar bands
put constraints on the interstellar reddening which is estimated
to be E(B$-$V)=$0\fm85\pm0\fm3$.
The available visual and CCD-V observations are used to determine
t$_{\rm 0}$,
t$_{\rm 2}$ and t$_{\rm 3}$. The resulting parameters are: 
t$_{\rm 0}$=2451669.5$\pm$0.1, t$_{\rm 2}=30\pm1$ d, t$_{\rm 3}=36\pm1$ d.
The recent lightcurve is found to be generally similar
to that observed in 1917 with departures as large as 1--2 mag in certain
phases. This behaviour is also typical for the U~Sco subclass.

 
\end{abstract}

\section{Introduction}

The first documented outburst of CI~Aql was discovered in 1917
at a photographic maximum of 11\fm0 (Reinmuth 1925).
The reported $\sim4\fm6$ eruption and the lack of subsequent
outbursts suggested CI~Aql to be a classical nova or dwarf nova with
long recurrence time. The dwarf nova classification was
disproved by spectroscopic observations which showed no
Balmer emission lines (Szkody \& Howell 1992). Very recently,
Wilson (2000) examined the Harvard College Observatory patrol plates
taken in 1917 and found the 1917 outburst much brighter than
Reinmuth's estimate. Tycho B magnitudes for nearby comparison
stars gave a maximum brightness of 8\fm6.

Time resolved CCD photometric observations by Mennickent \&
Honeycutt (1995) lead to the discovery of eclipsing binary
nature of CI~Aql with an orbital period of 0.618355 days.
Their subsequent optical and far red spectroscopy
showed only absorption lines of hydrogen, sodium and calcium.
The Ca II infrared triplet being a considerably good
luminosity indicator implied an evolved giant
secondary component and the authors suggested CI~Aql
to belong to the class of recurrent novae. However,
the lack of short-term erratic light variation (flickering)
lead to a conclusion that the components of
CI~Aql were not in strong interaction between 1991-1995.
In contrast to above, Greiner et al. (1996) presented
optical spectra in the 3800--7200 \AA\ range showing
the high-excitation lines HeII $\lambda4686$ and
NIII/CIII $\lambda4650$ in emission. Also, they
found the [O I] emission at 6300 \AA. Based on a simple
consideration about the mass dependence of the
semi-major axis for a {\it reasonable} range of masses
of the two components, Greiner et al. excluded the
possiblity of a giant component in the system, but
suggesting subgiant or main-sequence components. However,
their results were inconclusive on the real nature of
CI~Aql. The present outburst cleared the situation,
placing CI~Aql among the recurrent novae.

Recurrent novae (RNe) form a rare class of cataclysmic variables.
Their repeated outbursts are thought to be caused by thermonuclear
runaways on the surface of white dwarf primaries which accrete
material from the cool dwarf or giant secondary stars. Three subclasses
(T~Pyx, U~Sco, and T~CrB) can be distinguished according
to the physical nature of the secondary and the corresponding
orbital period (Warner 1995). The 0.618 d orbital period of
CI~Aql is typical for the U~Sco subclass (V394~CrA, LMC-RN, U~Sco)
consisting of RNe with He-dominated quiescent disks,
$P_{\rm orb}\sim 1$ d and visible secondary spectra in quiescence.
The declining branches of their lightcurves
vary from outburst to outburst indicating intrinsic variations
of the interaction between the components.

The 2000 outburst of CI~Aql
was discovered by K. Takamizawa on unfiltered T-Max
400 exposures taken on April 28.669 and 28.673 UT. An independent
photographic discovery was reported by M. Yamamoto.
The first low-resolution spectra (April 29.6 UT) showed prominent
H$\alpha$ emission with no P-Cyg profile indicating a fast nova
at a slightly evolved stage (Takamizawa et al. 2000). Precise
astrometry by Yamaoka et al. (2000) resulted in the identification
of the star with CI~Aql known as an eclipsing binary
(Mennickent \& Honeycutt 1995) and Nova~Aql 1917 (Reinmuth 1925,
Duerbeck 1987). Their coincidence was provided by the large scale
of the discovery plates taken in 1917.
Objective prism spectra by W. Liller showed
strong H$\alpha$ emission, weak H$\beta$ line, [O I] lines
(6300 and 6360 \AA) and He II at 5870 \AA (Yamaoka et al. 2000).
Early multicolour photometric data obtained around maximum
(Yamaoka et al. 2000, Wilson et al. 2000, Hanzl 2000) gave
(B$-$V)$_{\rm max}\approx0\fm7-0\fm8$ suggesting relatively high
reddening, also pointed out by Liller based on the observed
weak H$\beta$ emission compared to the strong H$\alpha$ line.
Mazuk et al. (2000) reported 0.8--2.5 micron spectrophotometry
obtained 74 days after the peak brightness, which showed
evidence for an increased excitation of the emission-line
gas, as the neutral C, N and O lines disappeared and the
prominent He II lines replaced the He I 10830 \AA\ line.

The main aim of this paper is to present optical spectra taken
around and after the maximum, between $\Delta$t = $-$0.6 d to
$\Delta$t = +53 d. The low- and medium resolution spectra
were used to determine the main outburst properties, the
expansion velocity and the interstellar
reddening. We also use all publicly available visual
photometric data collected by
the VSNET group ({\tt http://www.kusastro.kyoto-u.ac.jp/vsnet})
to estime the rates of decline and to check the photometric
phases of the obtained spectra.

The paper is
organised as follows. Observations and data reductions are
briefly described in Sect.\ 2, the detailed description
of the obtained spectra is given in Sect.\ 3, while
the reddening determination is discussed in Sect.\ 4.
We present the visual lightcurve in Sect.\ 5 and compare
it with the 1917 outburst. Concluding remarks are
given in Sect.\ 6.

\section{Observations}

\begin{table*}
\begin{center}
\caption{Journal of spectroscopic observations}
\begin{tabular} {llllll}
\hline
Date (UT)  &  Hel. JD        & Instrument & dispersion & range (\AA) & $\Delta$t\\
2000           &  (2451000+) &            & element    &             & (day) \\
\hline
May 3/4    &  668.867        & 1.88-m &  1800 l/mm     & 5080--5290  & $-$0.6\\
           &  668.875        & 1.88-m &  1800 l/mm     & 6500--6700  & $-$0.6\\
May 5/6    &  670.884        & 1.88-m &  1800 l/mm     & 6500--6700  & +1.3\\
May 8/9    &  673.741        & 1.88-m &  1800 l/mm  & 5080--5290  & +4.2\\
           &  673.755        & 1.88-m &  1800 l/mm  & 6500--6700  & +4.2\\
May 10/11  &  675.860        & 1.88-m &  1800 l/mm  & 5080--5290  & +6\\
           &  675.876        & 1.88-m &  1800 l/mm  & 6500--6700  & +6\\
May 15/16  &  680.818        & 1.88-m &  600 l/mm  &  3860--4490  & +11\\
           &  680.837        & 1.88-m &  1800 l/mm &  6500--6700  & +11\\
May 25/26  &  690.644        & 1.88-m &  1800 l/mm &  6500--6700  & +21\\
           &  690.867        & 1.88-m &  1800 l/mm &  5800--6000 & +21\\
May 26/27  &  691.452        & 0.6-m Schmidt & 5$^\circ$ prism  &  4200--8900 & +22\\
May 27/28  &  692.453        & 0.6-m Schmidt & 5$^\circ$ prism  &       4200--8900 & +23\\
May 28/29  &  693.472        & 0.6-m Schmidt & 5$^\circ$ prism  &       4200--8900 & +24\\
May 29/30  &  694.415        & 0.6-m Schmidt & 5$^\circ$ prism  &       4200--8900 & +25\\
May 30/31  &  695.419        & 0.6-m Schmidt & 5$^\circ$ prism  &       4200--8900 & +26\\
June 6/7   &  702.711--702.839 & 1.88-m & 1800 l/mm & 6500--6700 & +33\\
           &  (8 spectra)      &        &           &            &    \\
June 21/22 &  717.802        & 1.88-m  &  1800 l/mm & 6500--6700 & +48\\
June 26/27 &  722.765        & 1.88-m  &  1800 l/mm & 6500--6700 & +53\\
\hline
\end{tabular}
\end{center}
\end{table*}

The presented data were aquired on 14 nights at two observatories in
May and June, 2000. Medium resolution ($\lambda/\Delta\lambda\approx7000-10000$)
spectroscopic observations
were carried out on 9 nights with the Cassegrain-spectrograph
attached to the 1.88-m telescope of the David Dunlap Observatory
(Richmond Hill, Canada). The detector was a Thomson 1024x1024
CCD chip (with a 6 e$^-$ readout noise).
The slit width was 306$\mu$ corresponding to 1\farcs8 on the sky.
Typical observing circumstances at DDO are far from being photometric which
is reflected, for instance, in the usual seeing values (2-3").
That is why we did not attempt to flux calibrate the data.
All spectra presented
throughout the paper were continuum normalized, though it was a quite
difficult task in certain cases. Further observational
details are given in Table\ 1.

The DDO spectra were reduced with standard IRAF tasks. Nightly master
biases were created by forming a median of the individual bias frames.
The bias subtraction was followed by flat-fielding with a similar nightly
master flat-field from five to seven individual flat-field images.
The aperture extraction and wavelength calibration was done
with the task {\it doslit}. The wavelength scale was determined
in each case with two FeAr spectral lamp exposures obtained
immediately before and after every stellar exposure. The comparison
spectral lines were identified with the web-based
Iron-Argon Spectral Atlas by Willmarth \& Cheselka
\footnote{\tt http://www.noao.edu/kpno/specatlas/fear/fear.html}.
The integration times ranged between 200-1200 seconds,
depending on the actual brightness, wavelength range and resolving power.
As well as the nova, we observed HD~177724 (rapidly rotating
A-type star) to identify telluric lines. This turned
out to be crucial when distinguishing sharp absorption
features in the late high-resolution H$\alpha$ profiles.
The continuum normalization was made by fitting
low-order Chebyshev-functions (with the IRAF task {\it contin})
to those parts of the individual spectra which were not or only slighty
affected by the broad line profiles.

Low-resolution objective prism spectra were obtained on
five nights in May, 2000 at Piszk\'estet\H{o} Station
of the Konkoly Observatory with the 60/90/180 cm Schmidt-telescope.
The detector was a Photometrics AT200 CCD camera
(1536x1024 pixels, KAF-1600 chip with UV-coating).
The objective prism has a refracting angle of
5$^\circ$ giving an image scale of 580 \AA/mm at H$\gamma$.
In the case of objective prism spectroscopy, the resolving power
is grossly affected by the seeing that smooths the detected
image along and perpendicular to the dispersion. The resolution
($\lambda/\Delta\lambda$) was estimated from the width of the
spectral images (typically $\sim$3 pixels) and the actual
\AA/pixel image scale which strongly depends on the wavelength range.
The resulting resolution values are 290 and 110 for the blue
and red end of the spectra, respectively.
Although the unfiltered observations covered the whole optical
region between 3800 and 9000 \AA, the useful spectral coverage
was a slightly narrower (4200--8900 \AA). The full journal
of observations is presented in Table\ 1.

The spectral extraction of the objective prism frames was
done with routines developed by the first author. Briefly, an
automatic detection of the spectral images is performed, which
results in the location of the spectra perpendicular and along
the dispersion. For the wavelength calibration two comparison-star
spectra were used, where unambiguous spectral features
provide good calibrator data (hydrogen Balmer-series in an
A-type star, molecular bands in a M-type star). The common
features (strong atmospheric lines at 7200 \AA, 7600 \AA\ and
8170 \AA) helped adjust the spectra to the same
wavelength scale. The residual scatter of calibrating spectral lines is typically
about 1 pixel, corresponding to 5--20 \AA\ depending on the spectral
region.
The next step was a relative flux-calibration
by dividing the extracted spectra with the spectral response
function of the instrument (determined from the observed and
absolute flux distribution of Vega taken from Gray 1992).
However, this calibration may suffer from large
systematic errors as Vega was observed at a significantly
different air-mass with the shortest exposure time available
(1 msec). Since neither the observing
conditions were extremely good (strong variations of the
seeing), nor could the possible shutter effects
be included in the reduction, we normalized the
low-resolution spectra to the continuum. This way we have lost
the possible hints of a red continuum (due to the cooler secondary
star) but the most important spectral features could be well
identified.

\section{Description of the spectra}

\begin{figure}
\begin{center}
\leavevmode
\psfig{figure=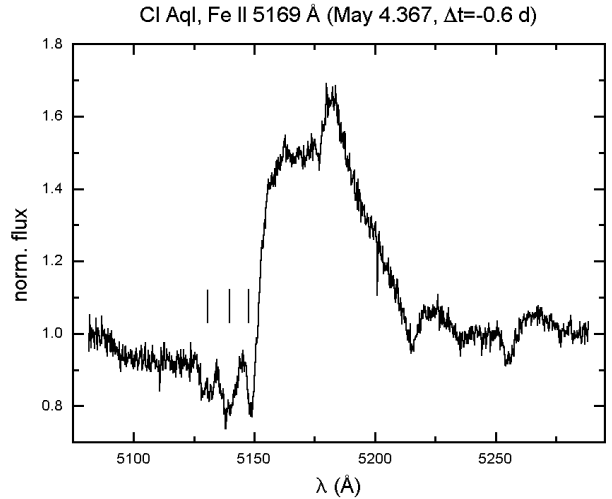,width=\linewidth}
\caption{The continuum normalized FeII 5169 \AA\ line before the light
maximum. Three absorption features shifted by $-$1100, $-$1700 and
$-$2200 km~s$^{-1}$ are marked.}
\end{center}
\label{f1}
\end{figure}

The first two spectra were obtained just before the light maximum, which
occured on May 5.0 UT (JD 2451669.5, see later). The bluer one
was centered at 5184 \AA\ and showed the strong Fe II 5169 \AA\
emission line with P-Cyg profile (Fig.\ 1).
The radial velocities of absorption features are $-$1100, $-$1700
and $-$2200 km~s$^{\rm -1}$. (This wavelength range was observed again
on two other nights, but the acquired spectra had too poor S/N ratios
for further conclusions).
The simultaneously observed H$\alpha$
line is nearly Gaussian
with slight redward asymmetry (top curve in Fig.\ 6). Note the
presence of the diffuse interstellar band (DIB) at 6613 \AA.
The later evolution of the H$\alpha$ line
will be described separately, thus we bring forward discussion
of other spectra.

\begin{figure}
\begin{center}
\leavevmode
\psfig{figure=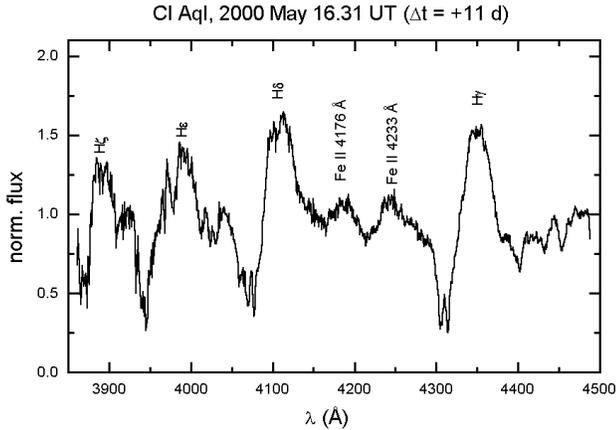,width=\linewidth}
\caption{The intermediate resolution spectrum of CI~Aql in the
blue region}
\end{center}
\label{f2}
\end{figure}

One intermediate resolution spectrum was taken on May 16.3 UT
centered at 4200 \AA\ ($\Delta$t = +11 d). Prominent Balmer emission
lines with strong P-Cyg profiles dominate the spectrum, while
Fe II 4176 and 4233 \AA\ lines are also present (Fig.\ 2).
An H$\alpha$ observation on the same night reveals the
similarity of all observed hydrogen lines: double-structured
absorption suggesting two expanding shells with velocities
of $-$1800 and $-$2400 km~s$^{\rm -1}$ (Fig.\ 3).

\begin{figure}
\begin{center}
\leavevmode
\psfig{figure=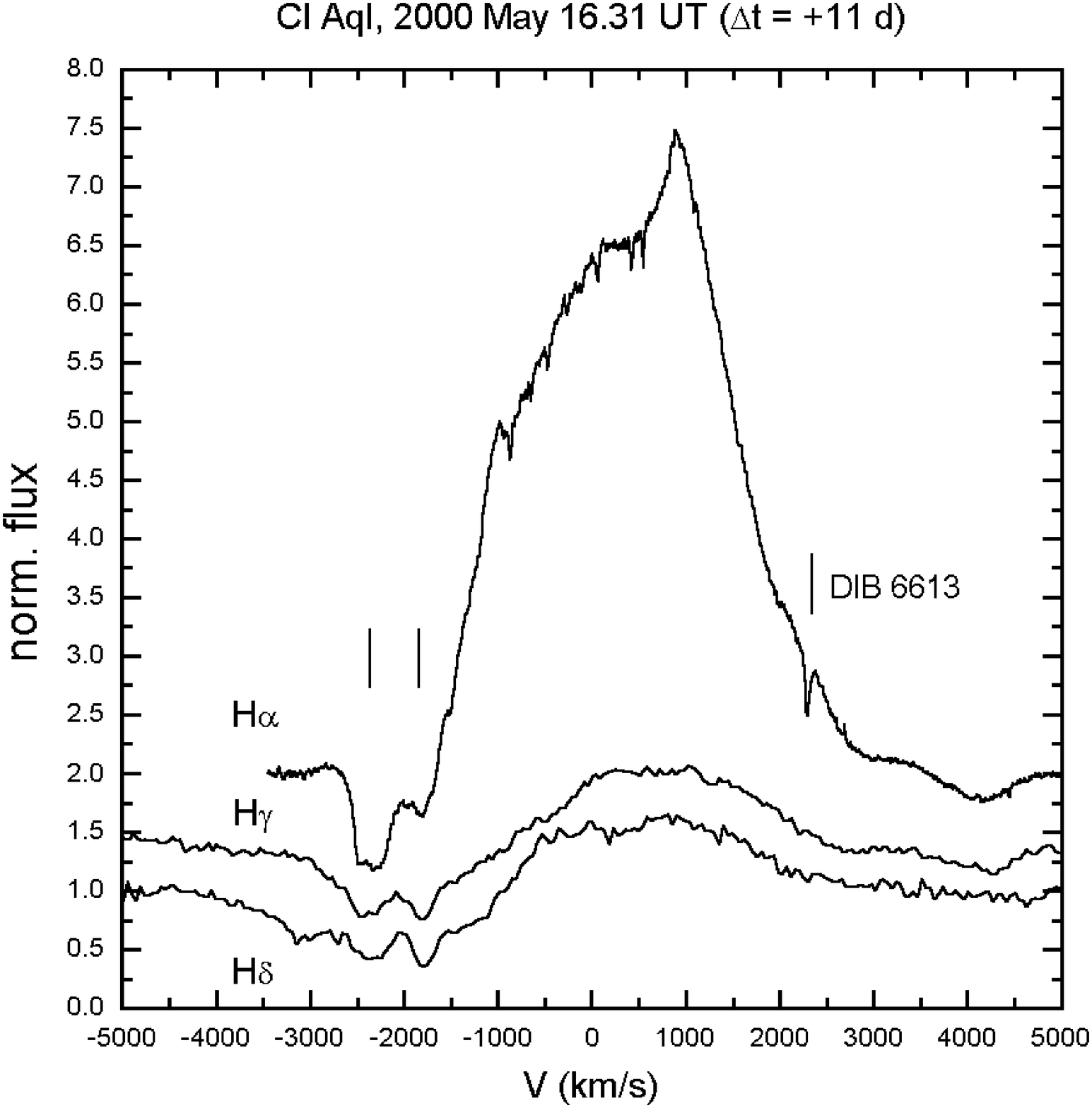,width=\linewidth}
\caption{Medium resolution hydrogen profiles. Two well-defined
absorptions are present at $-$1800 and $-$2400 km~s$^{-1}$. The
broad windgs extend to $\pm$3000--4000 km~s$^{-1}$. The continuum
normalized spectra were shifted by 0.5 for clarity.}
\end{center}
\label{f3}
\end{figure}

\begin{figure}
\begin{center}
\leavevmode
\psfig{figure=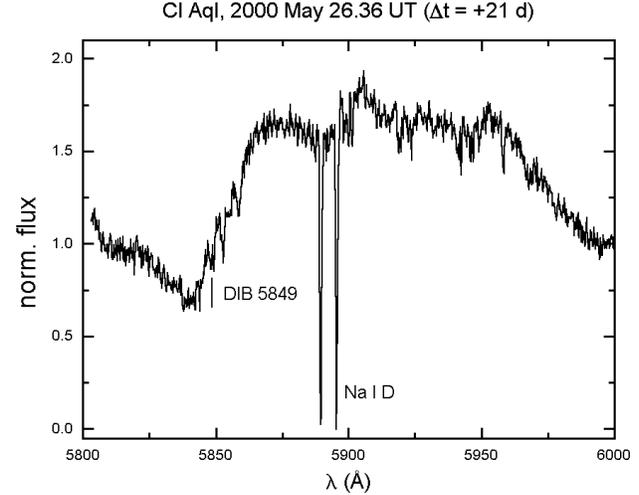,width=\linewidth}
\caption{The He I 5876 and Na I D blend.
Note the presence of a strong interstellar component
implying fairly high reddening}
\end{center}
\label{f4}
\end{figure}

The next run was on May 25 ($\Delta$t = +21 d),
when the He~I~5876 \AA/Na~I~D
blend was observed. The most important detection in this
spectrum is of the strong interstellar component
(see the deep and sharp features at the rest wavelengths
of sodium doublet in Fig.\ 4) and the much weaker DIB at 5849 \AA.
These lines can be used as reddening indicators as will be discussed
in the next section.

\begin{figure}
\begin{center}
\leavevmode
\psfig{figure=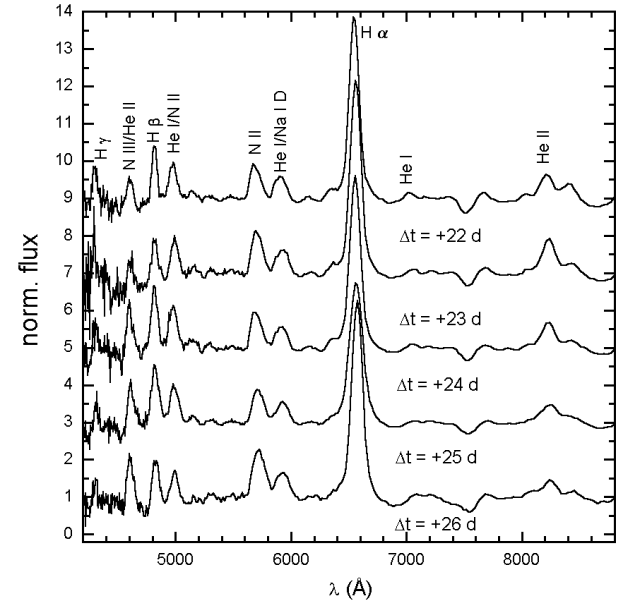,width=\linewidth}
\caption{A series of low-resolution objective prism spectra for CI~Aql.
A vertically shift of 2 was applied to avoid overplotting}
\end{center}
\label{f5}
\end{figure}

The first low-resolution objective prism spectrum taken
a day later is shown in Fig.\ 5. We could identify the
following lines/blends: hydrogen Balmer series from
H$\alpha$ to H$\gamma$, N III 4640/He II 4686,
N III 5001/He I 5016, N II 5679, He I 5876/Na I D, He I 7075,
He I 7281 and He II 8237 \AA.  Further objective prism spectra taken in the
next four days did not show significant variations, except
some changes in the He II lines, while the N III/He II blend became
stronger than H$\beta$. All of these features are typical
for a ``He/N'' nova during the permitted phase (see Fig.\ 2 in
Williams 1992). Additionally, we took a 3-hours long time-series
objective prism observation on May 30, but did not find
significant changes.

Finally, the largest number of spectra addressed the evolution
of the H$\alpha$ line. Being the strongest emission line,
it could be observed even in the fainter state in late June.
Fig.\ 6 summarizes the line profile variations. The early
quasi-Gaussian profile changed to a P-Cyg profile implying
expansion velocities above 2000 km~s$^{-1}$. The broad wings
extend to $\pm$3000-4000 km~s$^{-1}$ suggesting an early FWZI
of $\sim$8000 km~s$^{-1}$. The detected 200 \AA\ wide region
prevented obtaining a proper FWZI-curve as has been presented for
U~Sco by Munari et al. (1999) and Anupama \& Dewangan (2000), but
parallel to the shape variations, the H$\alpha$ line became slightly
narrower. The strongest line profile
change occured between day +21 and +33, when the saddle-shaped
profile formed with two maxima at $\pm$1100 km~s$^{-1}$.

We have also tried to find short-term variations in the
line profile by taking 8 spectra in a 3-hours long period
on June 6, but the attempt has failed.

The presented spectra will be compared with other RNe in Sect.\ 6, here
we only mention that the observed behavior is very similar to
other U~Sco-type RNe (U~Sco, V394~CrA), only the time scales are
different.

\begin{figure}
\begin{center}
\leavevmode
\psfig{figure=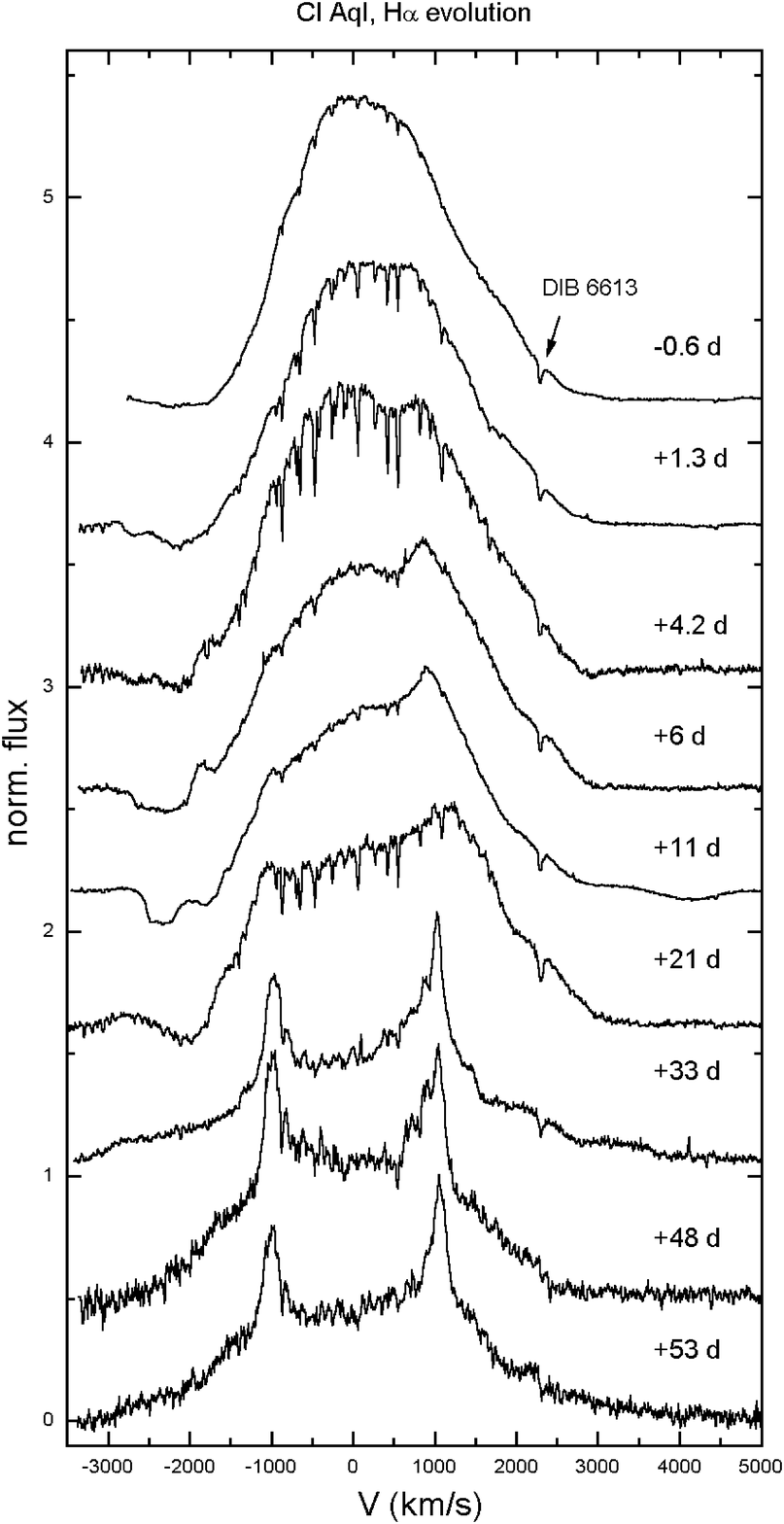,width=\linewidth}
\caption{The evolution of the H$\alpha$ profile during the
first two months after the maximum. The continuum
spectra were re-normalized at $\lambda$=6586 \AA\
(V=1100 km~s$^{-1}$) in order to
enable an easy comparison of the line shapes (the continuum
normalized flux maxima ranged between 6 and 150). Each
subsequent spectrum is shifted upwards by 0.5}
\end{center}
\label{f6}
\end{figure}

\section{The interstellar reddening}

As has been noted firstly by Liller (2000), CI~Aql suffers
from a significant amount of interstellar reddening. Since the most
interesting parameters of a nova system depends criticially on the
inferred luminosity, and consequently, the distance, an
accurate determination of the colour excess and visual extinction
is highly desirable.
Several spectroscopic methods using absorption features
originating from the interstellar matter in the line of sight
exist (see, e.g. Jenniskens \& D\'esert 1994,
Munari \& Zwitter 1997, Oudmaijer et al. 1997), though these
methods suffer from significant limitations. For instance, the Na I D
doublet provides reliable excesses only in the moderately
reddened region (up to E(B$-$V)=0\fm4, Munari \& Zwitter 1997).
Some of the diffuse interstellar bands (DIBs) provide good reddenings
(e.g. DIB 5849, Oudmaijer et al. 1997), while other give only rough
estimates. The internal consistency is in the order of 0\fm1--0\fm2 even
in the best cases, therefore, a certain amount of uncertainty
cannot be exceeded.

We have surveyed all of the medium resolution spectra to identify
possible DIBs taken from the list of Jenniskens \& D\'esert (1994).
We have unambiguously find DIB 5849 and 6613. Their equivalent
widths ($W$) was measured using the IRAF task {\it splot}. The resulting
values are: $W_{\rm 5849}=0.04\pm0.01$ \AA\ and
$W_{\rm 6613}=0.25\pm0.03$ \AA.
Unfortunately, the latter is slightly affected by a telluric line at
6612 \AA, that is why its width has a larger uncertainty.
Jenniskens \& D\'esert (1994) gave the following ratios for
the $W$/E(B$-$V): 0.048 (err. 0.008) for 5849 and 0.231 (err. 0.037) for
6613. The corresponding reddenings are E(B$-$V)$_{\rm 5849}=0\fm83\pm0\fm20$
and E(B$-$V)$_{\rm 6613}=1\fm08\pm0\fm20$.
The interstellar line Ca II 3933.66
can be also used through the empirical relationships between the width of
DIB 5780 and interstellar lines (see Table\ 2 in Jenniskens \& D\'esert
1994). Therefore, although we have not detected DIB 5780, we could
convert the measured Ca II equivalent width to $W_{\rm 5780}$ which
resulted in an E(B$-$V)=0\fm66$\pm$0\fm30.

In the case of CI~Aql the Na~I~D doublet is of lower significance due to
the saturation effects. Interestingly, as has been noted by the
referee, there is an apparent difference between the strength of Na~I~D
lines obtained by us and that of Greiner et al. (1996), i.e. the latter
data suggest a weaker, unresolved doublet. A real difference would
query the interstellar origin of this resonance line and the whole reddening
estimation should be reconsidered. We attribute this
phenomenon to the lower resolution of that spectrum by Greiner et al.
(1996), because a close inspection of their Fig.\ 1 reveals a broad
($\sim$15 \AA) single Na~I~D line. We could reproduce this kind of
appearance with a resampling and Gaussian convolution of our spectrum
mimicking the same resolution as quoted by Greiner et al.
(about 1 \AA\ FWHM). However, we cannot solidly exclude the
possibility of other origin, e.g. some kind of circumstellar absorption.
The measured equivalent widths are
$W_{\rm Na D1}=0.84\pm0.02$ \AA\ and $W_{\rm Na D2}=0.76\pm0.02$ \AA.
Their ratio is 1.10, far from the theoretically expected 2.0 at the
lowest optical depths, but exactly what is found for the asymptotic
behaviour at high reddenings (Munari \& Zwitter 1997).
If one checks the relation between equivalent
width and reddening presented in Fig.\ 2 in Munari \& Zwitter (1997),
only a weak and approximative conclusion can be drawn as
0\fm8$<$E(B$-$V)$<$1\fm5.

Further constraints on the reddening are provided by the published colour
measurements. Hanzl (2000) gave B$-$V=$0\fm69\pm0\fm02$ on May 7.03 UT
($\Delta$t = +2 d), while Jesacher et al. (in Wilson et al. 2000) presented
B$-$V=0\fm82 on May 10.97 UT ($\Delta$t = +6 d). The B$-$V colour
of novae around maximum tends to be about B$-$V=$0\fm23\pm0\fm06$ with a
significant dispersion of $\sigma=0\fm16$ (Warner 1995).
The resulting reddening lies between 0\fm46--0\fm59 mag (with 0\fm06 formal
error). Two magnitudes down from maximum the dispersion decreases, therefore
the relation (B$-$V)$_{\rm 0}^{\rm V(max)+2}\approx$0\fm0 can be also
used. There are a few BV CCD photometric measurements in the
VSNET database obtained in early June, resulting in again a reddening
about 0\fm5. However, three prominent emission complexes are
covered by the standard V passband (N II 5679, He I 5876/Na I D, H$\alpha$),
therefore we consider these colour measurements to be heavily affected
by the presence of such strong emission.
It is difficult to say which reddening is more reliable. In the
following discussion we adopt the unweighted mean of spectroscopic
values which is 0\fm85$\pm$0\fm3 (formal error). Despite
the limitations of using interstellar lines, their observational
data are much less affected by the nova itself as in the case of
multicolour photometry.

\section{The light curve}

\begin{figure*}
\begin{center}
\leavevmode
\psfig{figure=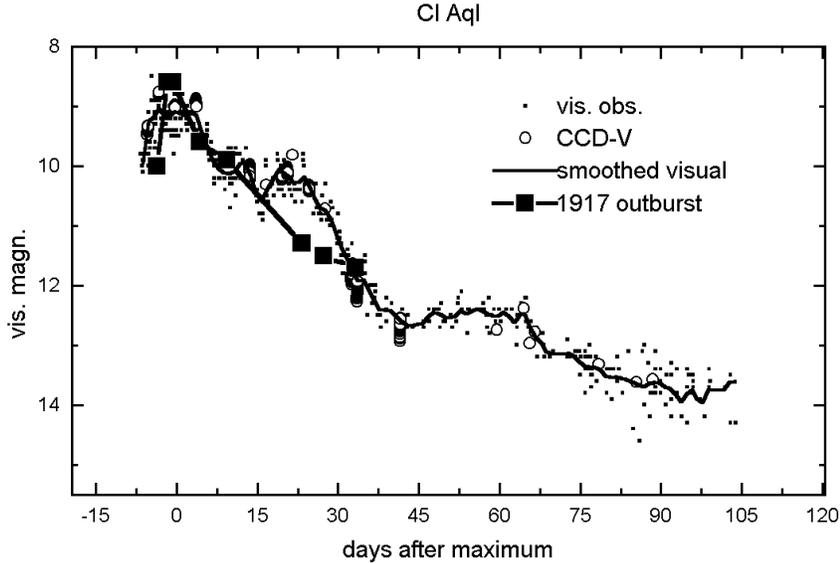,width=12cm}
\caption{The visual lightcurve of the outburst. The light curve
of the 1917 outburst determined by Williams (2000) is shown for
comparison. The maximum occured on May 5.0, 2000 (JD 2451669.5)}
\end{center}
\label{f7}
\end{figure*}

In order to estimate the light curve parameters (epoch of maximum,
rates of decline), we used the CI~Aql record in the VSNET archive
(this data also includes the observations published in the IAU
Circulars). It contains 2309 individual data points between
April 28 and August 16, 2000 (1737 CCD observations and
572 visual estimates). We plot the light curve in Fig.\ 7.
The epoch of maximum was determined by fitting a low-order polynomial
around the top of the light curve, while t$_{\rm 2}$ and t$_{\rm 3}$
were read off from the smoothed light curve (see Fig.\ 7).
The derived parameters
are: t$_{\rm 0}=2451669.5\pm0.1$ (2000 May 5.0 UT), t$_{\rm 2}=30\pm1$
days, t$_{\rm 3}=36\pm1$ days. Consequently, CI~Aql is a
moderately fast nova. The apparent magnitudes in maximum (9\fm0) and
minimum (16 mag) give a considerably low outburst amplitude
of $\sim$7 mag.
We plotted also the light curve of the 1917 outburst determined
by Williams (2000) in Fig.\ 7. The similarity is evident, though
a significant deviation is suggested at $\Delta$t = +20 d, when
an additional brightening occured in 2000. The maxima
were at similar brightness, but later differences as high as 2 mag
can be found. Recent t$_{\rm 2}$ (30 d) differs significantly
from that of in 1917 (18 d) which is entirely due to a plateau
of the light curve of the 2000 outburst. Consequently, t$_{\rm 3}$
appears to be more reliable indicator of the rate of decline.

Three maximum magnitude versus rate of decline (MMRD) relations
were used to calculate visual absolute magnitude (Della Valle \& Livio 1995,
Capaccioli et al. 1989 and Schmidt 1957). They result in
$-$7\fm4, $-$7\fm3 and $-$7\fm6. The constant absolute
magnitude 15 days after the maximum (Capaccioli et al. 1989)
gives $-$7\fm1. Their simple mean is M$_{\rm V}=-7\fm35\pm0\fm2$ (formal
error). However, the true uncertainty could be much larger as even the
applicability of MMRD relations for the RN outburst can be
questionned (e.g. the outbursts of U~Sco are regularly
underluminous compared to those of classical novae, Munari et al. 1999).
Any meaningful luminosity value would need proper modelling of either
the outburst or observations carried out in the quiescence, which is beyond
the scope of this paper.

\section{Discussion}

The presented spectroscopic behaviour largely resembles
other RNe in different phases. In the early phases of
outburst broad emission lines dominated the spectra. The
extensive wings reached FWZI$\sim$8000-9000 km~s$^{-1}$ (Fig.\ 1
and Fig.\ 3). Later they became narrower. The low-resolution
spectra covering the whole optical range show broad
emission complexes $\lambda\lambda$4500--4700,
$\lambda\lambda$4900--5100, $\lambda\lambda$5600--5800,
$\lambda\lambda$5800--6000. Very similar spectra
were obtained, e.g., for V394~CrA 5 and 6 days after maximum (Sekiguchi
et al. 1989a, Williams et al. 1991, Williams 1992) and
for U~Sco at +1.45 days (Anupama \& Dewangan 2000)
and between +3 and +5 days (Munari et al. 1999). Interestingly,
Munari et al. (1999) observed exactly the same change
of He II lines at 8237 \AA\ and 4686 \AA\ in U~Sco between
+1.63 to +4.59 days (t$_2$=2.2 days, t$_3$=4.3 days)
as has been found by us between +22 and +26 days (t$_2$=30 days,
t$_3$=36 days). This fact suggests that the underlying physical
mechanisms are similar.

The general
appearance is typical for ``He/N'' spectra defined by
Williams (1992). This gives further support to the statement
of Williams (1992), that almost all novae with short
recurrence times appear to have ``He/N'' type spectra
(U~Sco, V394~CrA, V745~Sco, V3890~Sgr, LMC 1990 No. 2).
The interpretation of this classification was given by
Williams (1992) in terms of different components in the
ejecta. The broader lines of the ``He/N'' spectrum originate
in a discrete shell, which is ejected at considerably high
velocities from the white dwarf surface at the peak
of outburst.

In our dataset there are some weak pieces of evidence for another
interpretation. The fairly strong and broad Fe II 5169 line around
the maximum raises the possibility that CI~Aql belongs to
the hybrid objects as discussed in Williams (1992). These stars
have ``Fe II'' type spectra in the early phases that evolve to
the ``He/N'' type. The very early ($\Delta$t = $-$7 d)
low-resolution spectrum taken by W. Liller showed no evidence
of Fe II lines (Yamaoka et al. 2000), while our spectrum
a week later showed strong and broad Fe II 5169 emission (Fig.\ 1).
Also, the intermediate resolution blue spectrum taken at +11 d (Fig.\ 2)
is very similar to what is observed in Nova LMC 1988 No. 2 by
Sekiguchi et al. (1989b). Williams (1992) suggested that
this behaviour is caused by two distinct phases of the outburst.
The ``Fe II'' spectra probably originate from discrete and massive
shells, which are optically thick, thus causing the early
photosphere to occur in the ejected shell. As the expanding
shell(s) becomes optically thin, the spectrum change to the
``He/N'' type with more rectangular line shapes. The double-structured
absorption in the P-Cyg profiles (Figs.\ 1 and 3) implies
a quite complex inner structure of the ejecta, thus may strengthen
the previous explanation.

The H$\alpha$ evolution is also similar to what has been found
in other RNe. Munari et al. (1999) presented a nice coverage
between +0.64 d and +22.6 d for U~Sco, in which the early
saddle-like H$\alpha$ split into three components with velocity
separation of the order of $\pm$1600 km~s$^{-1}$. While the three
components in the eclipsing system of U~Sco are difficult to
interpret in terms of collimated beams of material ejected
at a large angle from the orbital plane, the two components
of CI~Aql can be explained with simple equatorial and polar
rings of enhanced brightness in the ejected shell (Gill \& O'Brien
1999). Since CI~Aql is an eclipsing system, we see the hypothetic
rings nearly edge-on. As has been pointed out by Gill \& O'Brien (1999),
the high inclination means that the least information on the
shell structure can be derived from the line profiles.

Finally, the main conclusions of this paper can be summarized as follows:

\begin{itemize}

\item Optical spectra taken between $-$0.6 and +53 d are presented and
discussed. The overall appearance is similar to other recurrent
novae, and consequently, the star belongs to the ``He/N'' type novae
defined by Williams (1992). This means the lines originate from
a discrete shell ejected by a velocity of 2000--2500 km~s$^{-1}$.
Weak evidence for possible changing the type from
``Fe II'' to ``He/N'' is present in our dataset.

\item We identified two diffuse interstellar bands which were
used to estimate the interstellar reddening. The saturated
interstellar Na I D doublet supports the fairly high
colour excess adopted to be E(B$-$V)=0\fm85$\pm$0\fm3.

\item The visual light curve was used to determine epoch of
maximum light and rates of decline. The recent light curve
was compared with that of in 1917 and significant differences
were found. t$_{\rm 3}$ appears to be a more reliable indicator
of the decline.

\end{itemize}

\begin{acknowledgements}
This research was supported by the ``Bolyai J\'anos'' Research
Scholarship of LLK from the Hungarian Academy of Sciences,
Hungarian OTKA Grant \#T032258 and Szeged Observatory Foundation.
The warm hospitality of the staff
of the Konkoly Observatory and their provision of telescope time
is gratefully acknowledged.
The NASA ADS Abstract Service was used to access data and references.
This research has made use of Simbad Database operated at CDS-Strasbourg,
France.
\end{acknowledgements}


\begin{thebibliography}{}

\bibitem[2000]{anup00}
    Anupama G.C., Dewangan G.C. 2000, AJ 119, 1359

\bibitem[1989]{capa89}
    Capaccioli M., Della Valle M., D'Onofrio M., Rosino L. 1989, AJ 97, 1622

\bibitem[1995]{della95}
    Della Valle M., Livio M. 1995, ApJ 452, 704

\bibitem[1987]{duer87}
    Duerbeck H.W. 1987, Space Sci. Rev. 45, 1

\bibitem[1999]{gill99}
    Gill C.D., O'Brien T.J. 1999, MNRAS 307, 677

\bibitem[1992]{gray92}
    Gray D.F. 1992, Observations and analysis of stellar
    photospheres, Cambridge University Press, New York

\bibitem[1996]{grein96}
    Greiner J., Alcala J.M., Wenzel W. 1996, IBVS No. 4338

\bibitem[2000]{hanzl00}
    Hanzl D. 2000, IAUC No. 7444, 3

\bibitem[1994]{jen94}
    Jenniskens P., D\'esert F.-X. 1994, A\&AS 106, 39


\bibitem[2000]{mazuk00}
    Mazuk S., Rudy R.J., Lynch D.K. et al., 2000, IAUC 7490, 2

\bibitem[1995]{menn95}
    Mennickent R.E., Honeycutt R.K. 1995, IBVS No. 4232

\bibitem[1997]{munari97}
    Munari U., Zwitter T. 1997, A\&A 318, 269

\bibitem[1999]{munari99}
    Munari U., Zwitter T., Tomov T. et al. 1999, A\&A 347, L39

\bibitem[1997]{oud97}
    Oudmaijer R.D., Busfield G., Drew J.E. 1997, MNRAS 291, 797

\bibitem[1925]{rein25}
    Reinmuth K. 1925, AN 225, 385

\bibitem[1957]{schmidt57}
    Schmidt T. 1957, Z. Astrophys. 41, 182

\bibitem[1989a]{seki89a}
    Sekiguchi K., Catchpole R.M., Fairall A.P. et al. 1989,
    MNRAS 236, 611

\bibitem[1989b]{seki89b}
    Sekiguchi K., Kilkenny D., Winkler H., Doyle J.G. 1989, MNRAS 241,
    827

\bibitem[1992]{szkody92}
    Szkody P., Howell S.B. 1992, ApJS 78, 537

\bibitem[2000]{tak00}
    Takamizawa K., Kato T., Yamamoto M. et al. 2000, IAUC 7409, 1

\bibitem[1995]{warner95}
    Warner B. 1995, Cataclysmic variable stars, Cambridge Univ.
    Press, Cambridge

\bibitem[1991]{will91}
    Williams R.E., Hamuy M., Phillips M.M. et al. 1991, ApJ 376, 721

\bibitem[1992]{will92}
    Williams R.E. 1992, AJ 104, 725

\bibitem[2000]{williams00}
    Williams D.B. 2000, IBVS No. 4904

\bibitem[2000]{wilson00}
    Wilson J.C., Dunscombe K.R., Jesacher M.O. et al. 2000,
    IAUC No. 7426, 2

\bibitem[2000]{yam00}
    Yamaoka H., Ayani K., Shirakami K. et al. 2000, IAUC No. 7411, 1

\end{thebibliography}
\end{document}